\documentstyle[aps,graphicx,pra]{revtex}

\def\be{\begin{equation}}
\def\bea{\begin{eqnarray}}
\def\bma{\begin{mathletters}}
\def\ee{\end{equation}}
\def\eea{\end{eqnarray}}
\def\ema{\end{mathletters}}

\begin{document}

\draft

\title{Creation of entangled states of distant atoms by interference}

\author{C. Cabrillo,$^{(1)}$ J. I. Cirac,$^{(2)}$ P. Garcia--Fernandez,$^{(1)}$ and 
P. Zoller$^{(2)}$}

\address{(1) Instituto de Estructura de la Materia, Serrano 123, 
E-28006, Madrid, Spain}

\address{(2) Institut f{\"u}r Theoretische Physik, 
Universit{\"a}t Innsbruck, A--6020 Innsbruck, AUSTRIA}

\maketitle

\begin{abstract}
We propose a scheme to create distant entangled atomic states. It
is based on driving two (or more) atoms with a weak laser pulse, so that
the probability that two atoms are excited is negligible. If the
subsequent spontaneous emission is detected, the entangled state is
created. We have developed a model to analyze the fidelity of the
resulting state as a function of the dimensions and location of the
detector, and the motional properties of the atoms. 
\end{abstract}


\date{\today}

\widetext

\section{Introduction}

The preparation of entangled atomic states is one of the goals of Atomic
Physics and Quantum Optics. These states are a key ingredient for
studying some fundamental issues of Quantum Mechanics \cite{1}, as well
as for certain applications related to Quantum Information \cite{2}.
Methods proposed so far to ``engineer'' entanglement between atoms in
the laboratory are based on achieving and controlling an effective
interaction between the atoms that are to be entangled. Typically, these
interactions are mediated by the electromagnetic field. For example, in
cavity QED, two atoms can be entangled if they both interact with the
same cavity mode \cite{3}. This coupling of the two atoms to the field
mode can be simultaneous or sequential (that is, one atom interacts
first with the cavity mode, and then the other one). With trapped ions,
entangled states can be produced by using the Coulomb repulsion between
the ions, together with some laser couplings \cite{4}. With these
methods, it is always necessary that the atoms interchange some
particles (photons) or that they are very close to each other.

In this paper we propose a scheme to prepare entangled atomic states
using a different approach. In particular, the entangled state is not
produced by an effective interaction between the atoms, but rather by an
interference effect and state projection accompanying a measurement.
Imagine that we have two atoms $A$ and $B$, situated in distant
locations, both in an excited state $|0\rangle$. These atoms may decay
to the state $|1\rangle$ due to spontaneous emission, producing one
photon. A detector is placed at half the way between the atoms. After
some time, if the detector clicks and we cannot distinguish from where
the detected photon came, we will have produced an entangled state
\be
\label{Bell}
|\Psi\rangle = \frac{1}{\sqrt{2}}
(|0\rangle_A|1\rangle_B + e^{i\phi}|1\rangle_A|0\rangle_B) \,,
\ee
where $\phi$ is a fixed phase.  Entanglement is then achieved as a consequence
of two facts: first, the impossibility to determine from the detection 
event which atom emitted the photon; second, the projection postulate in Quantum
Mechanics, which indicates that after the detection the state of the
atoms is projected onto the one which is compatible with the outcome of
the measurement. The first effect is precisely the one that would give
rise to interference fringes at the detector position if one would
repeat several times the experiment, as it has been shown by the 
NIST group at Boulder \cite{5,6}. The second effect has been used, for example, in the
preparation of non--classical states of a cavity mode \cite{7}. Using this method
to prepare entangled states, the atoms do not need to interact, and no
interchange of particles (photons) is required. In fact, the
entanglement can be produced (in principle) in a time which is half the
distance between the atoms divided by the speed of light. 

In practice, the method described above might not be very useful. First,
it is very unlikely that the photon emitted by one of the atoms is
detected. Second, and more important, even if one photon is detected,
the second atom will eventually decay to the ground state thus yielding
the state $|1\rangle_A|1\rangle_B$, which is not entangled. Here we will
analyze in some detail how an experiment can be performed in a realistic
set--up. The idea is to use two three--level atoms with a Lambda
configuration (see Fig.\ \ref{fg:setup}). The states $|0\rangle$ and
$|1\rangle$ are the two ground states, so that once the state (\ref{Bell}) is
prepared, it will stay. Both atoms are initially prepared in the state
$|0\rangle$. The excitation is achieved by using a very short laser
pulse, which (with a small probability) excites one of the two atoms to
level $|2\rangle$. If following the excitation a
spontaneously emitted photon is detected, an entangled state of the two
atoms will be produced. The method presented here seems particularly
timely, in view of the spectacular experimental progress reported by the
NIST group of observation of interference fringes of the light emitted
by two independent atoms \cite{5}. In fact, the same experimental
setup could be used to prepare atomic entangled states using our
proposal.

In order to estimate the conditions that must be fulfilled to create an
entangled state, we have developed a theoretical model describing the
whole process of laser excitation of the two atoms, spontaneous emission
of a photon, and detection. The idea is to represent the detector as a
collection of atoms, and then to use master equation methods to describe
the projection occurring when a detection event is recorded. In this
way, the electromagnetic field does not appear explicitly in the
formulae, making the calculations simpler. We emphasize that the model
is equivalent to the one in which the whole state of the electromagnetic
field is taken into account at all times, and the measurement projects
its state along with the state of the atoms. This model can be easily
generalized to other situations in which there are more atoms present,
yielding entangled states of more than two atoms.

The paper is organized as follows: in Section II we explain
qualitatively the details of our proposal and discuss the main results,
and some of the practical problems. In Section III we present
the theoretical model. In Section IV we obtain an analytical formula
for the fidelity of the final state as a function of the physical
parameters involved in the problem. Finally, in Section IV we discuss
the results and point out some possible generalizations.

\section{Qualitative description}

Let us consider two atoms $A$ and $B$ separated by a distance $2d$. Each
of the atoms has an internal structure which can be described in terms
of a three--level Lambda system (see Fig.\ \ref{fg:setup}). It consists of two
ground levels $|0\rangle$ and $|1\rangle$, and an excited state
$|2\rangle$. A photodetector is located at a distance $D$ from the
segment connecting atoms $A$ and $B$ (see Fig.\ \ref{fg:setup}). The detector is
sensitive to photons of wavelength $\lambda_1$ (and/or polarization)
corresponding to the transition $|2\rangle\rightarrow|1\rangle$, which
is characterized by a spontaneous emission rate $\Gamma_1$. It is not,
however, sensitive to the ones corresponding to the other transition. 

Both atoms are initially prepared in the state $|0\rangle$. Then, they
are driven by a very short laser pulse on resonance with the transition
$|2\rangle\leftrightarrow|0\rangle$. As a consequence, sometimes one of
the atoms (or both) will spontaneously emit a photon of wavelength
$\lambda_1$, which might be recorded at the photodetector. Most of the times,
no photon will be detected after a waiting time $t\gg \Gamma_1$. In such
a case, the atoms are pumped back to the original state $|0\rangle$, and
the experiment is repeated until the detector clicks. Once this occurs,
the state of both atoms will be described by a density operator
$\rho_{A,B}$. The goal is to obtain a state as close as
possible to the maximally entangled state ({\ref{Bell}) where $\phi$ is a
phase that does not change from experiment to experiment. That is, we
wish to obtain a fidelity
\be
\label{fidel}
F=\langle\Psi|\rho_{A,B}|\Psi\rangle,
\ee
close to one. 

The physical idea is that the laser pulse prepares a superposition state
of the two atoms, which apart from the state $|0\rangle_A|0\rangle_B$
also contains a coherent superposition of the states
$|0\rangle_A|2\rangle_B$ and $|2\rangle_A|0\rangle_B$. Detection of a
photon implies that a transition $|2\rangle\rightarrow|1\rangle$ has
taken place in one of the atoms, producing a photon of wavelength
$\lambda_1$ that is detected. The term $|0\rangle_A|0\rangle_B$ will
thus be projected out from the atomic state, since it is incompatible
with that event (the state $|1\rangle$ of one of the atoms must be
present in the atomic state). Moreover, given the fact that the detector
cannot distinguish among photons emitted by different atoms, the
superposition of the states $|0\rangle_A|2\rangle_B$ and
$|2\rangle_A|0\rangle_B$ will be transformed into a superposition of the
states $|0\rangle_A|1\rangle_B$ and $|1\rangle_A|0\rangle_B$, i.e., it
will be close to the entangled state (\ref{Bell}).

In order to obtain an entangled state close to the ideal Bell state
(\ref{Bell}), several conditions have to be satisfied: (i) First, the
laser pulse has to be such that the probability of exciting both atoms
to the state $|2\rangle_A|2\rangle_B$ has to be much smaller than the
probability of exciting the relevant coherent superposition. Otherwise,
it may happen that although we detect a photon emitted by one of the
atoms, the other atom also emits a photon albeit in another direction
which is not detected; this would spoil the fidelity $F$ since the final
state of this process would be $|1\rangle_A|1\rangle_B$. In order to
avoid this problem one must use a sufficiently weak or short laser
pulse. In that case, the probability of exciting two atoms $\epsilon^2$
is of the order of the square of the probability of exciting only one
atom $\simeq 2\epsilon$. By choosing $\epsilon\ll 1$ one avoids the
two--atom excitation. Notice, however, that the laser beam cannot be too
weak since it would take a very long time to detect one spontaneously
emitted photon, given that the detection probability is proportional to
$\epsilon$. (ii) Second, the detector has to be sufficiently small. At
each point of the detector the phase $\phi$ will have a different value
spoiling the fidelity since a detection does not specify the exact
location of the event, and therefore the exact phase is unknown. Thus,
the detector has to be such that at all points the phase is practically
the same. In order to estimate the required size of the detector surface
one can use the analogy between the situation considered here and the
double slit experiment: the distance traveled by a photon coming from
one atom or the other will be somewhat different at different positions,
and therefore the accumulated phase depends on the position in which it
is detected. The phase will be essentially constant over regions where
the corresponding interference fringes have a constant visibility. Thus,
the length $L_x$ of the detector along the $XZ$--plane has to be much
smaller than the interfringe distance, $L_x\ll \lambda_1 D/d$. However,
the detection probability is proportional to the size of the detector
and therefore we cannot take $L_x$ arbitrarily small. (iii) Furthermore,
the dynamics of the atoms during the absorption emission cycle will also
affect the final fidelity. In fact, every absorption or emission of
photons by an atom is always accompanied by a recoil, which changes the
atomic motional state. This leaves a trace of which atom has emitted the
photon, thus also destroying the entanglement. In order to avoid this
problem, one has to find a way ``not to leave information about the
motional states behind.'' This can be done, for example, by using
trapped particles and operating in the Lamb--Dicke limit, where the
recoil energy does not suffice to change the atomic motional state
(similar to the M{\"o}sbauer effect). However, the extent to which this
effect can be reduced will also depend on the temperature of the atoms
in the trap, as well as on the propagation directions of the laser
beams.

In the following Sections we will solve in detail a theoretical model to
answer all of these questions. Our result is a simple
formula for the fidelity in which these effects are clearly separated.
We consider a situation where the atoms are trapped in identical isotropic
harmonic potentials, characterized by a frequency $\nu$ and 
initial temperature $T$. We obtain
\be
\label{Fid}
F=\frac{\cos^2(\theta_{\rm las})}{2} (1+ F_{\rm geo} F_{\rm dyn} )
\ee
where $\theta_{\rm las}$ is the pulse area (Rabi frequency times time),
and $F_{\rm dyn}$ and $G_{\rm geo}$ represent a dynamical and a
geometrical factor, respectively. More specifically, 
\be
\label{Fgeo}
F_{\rm geo} = {\rm sinc} \left[ \frac{dL_x}{2\lambda_1\sqrt{d^2+D^2}} \right],
\ee
where ${\rm sinc}(x)=\sin(x)/x$. We also have
\bea
\label{Fdyn}
F_{\rm dyn} &=& \int_0^\infty d\tau e^{-\tau} \exp\left\{
-2\eta^2 \coth\left(\frac{\hbar\nu}{2k_BT}\right) \times \right. \nonumber\\
&& \left.
\left[1-\cos(\chi)\cos\left(\frac{\nu\tau}{\Gamma}\right)
\right]\right\}.
\eea
Here, $\eta=2\pi a_{\rm tp}/\lambda_1$ is the so--called Lamb--Dicke parameter,
with $a_{\rm tp}=\sqrt{\hbar/2m\nu}$ the size of the harmonic trapping potential ground
state, $\Gamma$ is the total spontaneous emission rate from level $|2\rangle$,
and $\chi$ is the angle between the propagation direction of the laser
acting on an atom and the line that connects the atom with the center
of the detector (we take this angle to be the same for atoms $A$ and
$B$). 

The first factor in Eq.\ (\ref{Fid}) accounts for the effects due to the
laser excitation. That is, when $\theta_{\rm las}$ increases, the
fidelity decreases due to the fact that both atoms may be simultaneously
excited. The geometrical factor is related to the size of the detector
with respect to the interfringe distance. For small detectors compared
with such a distance, this factor approaches one. Finally, the dynamical
factor shows that the fidelity increases for small Lamb--Dicke
parameters and low temperatures, and depends on the ratio $\nu/\Gamma$
as well as the direction of the lasers. The highest fidelity occurs for
$\cos(\chi)\simeq 1$ and $\eta^2 \coth(\hbar\nu/2k_BT)
\ll(\Gamma/\nu)^2$. The first condition means that the laser direction
and the direction of the photon emitted and recorded at the detector has
to be practically the same. In that case the recoil given by the laser
is compensated by the recoil experienced by the atom in the spontaneous
emission process that is monitored at the photodetector, and therefore
no trace of which atom has emitted is left behind. Under such circumstances
a $\nu\ll\Gamma$ (weak confinement) is needed so that the atom does not
have time to oscillate in the trap before the spontaneous emission takes
place -- this would destroy the compensation of the recoils between the
absorption-emission process. In these limits we can approximate
\bma
\label{Factors}
\bea
F_{\rm geo} &\simeq& 1 - \frac{1}{6} 
  \left[\frac{dL_x}{2\lambda_1\sqrt{d^2+D^2}}\right]^2,\\
\label{eq:Fapro}
F_{\rm dyn} &\simeq& 1 - 
  2\eta^2 \coth\left(\frac{\hbar\nu}{2k_BT}\right) 
  \left(\frac{\nu}{\Gamma}\right)^2.
\eea
\ema

\noindent
On the other hand, under conditions of strong confinement ($\Gamma \ll \nu$) 
although it is not possible to compensate for the harmful effect of the recoil by
choosing the laser propagation direction, the dynamical factor can be very
close to one in the Lamb--Dicke limit ($\eta \ll 1$). In particular, for 
$\eta^2 \coth(\hbar\nu/2k_BT) \ll 1$ we have
\be
F_{{\rm dyn}} \simeq \;
1- 2\eta^2 \coth\left(\frac{\hbar\nu}{2k_BT}\right).
\ee

\section{Model}

\subsection{Master equation for the atoms and photodetector}

We consider two identical atoms $A$ and $B$, centered at positions
${\mathbf r}^A_0$ and ${\mathbf r}^B_0$, separated by a distance
$2d=|{\mathbf r}^A_0-{\mathbf r}^B_0|$. Each of the atoms has an
internal structure which can be described in terms of a three--level
Lambda system (see Fig.\ \ref{fg:setup}). It consists of two ground
levels $|0\rangle$ and $|1\rangle$, and an excited state $|2\rangle$.
Spontaneous emission from level $|2\rangle$ to both ground levels is
possible, and is characterized by the rates $\Gamma_{0,1}$ and
wavevectors ${\mathbf k}_{0,1}(\Omega)$, where $\Omega$ represents a
direction and $\Gamma=\Gamma_0+\Gamma_1$ the total decay width of the
excited state.

A detector of surface dimensions $S=L_xL_y$ and efficiency $\eta_D$ is
situated in the $XY$--plane, at a distance $D$ from the segment
connecting atoms $A$ and $B$. The center of the detector ${\mathbf r}_0$
and the center of atoms $A$ and $B$ define the $XZ$ plane. We will
describe the detector as a collection of independent point atoms located
at position ${\mathbf r}$, with ${\mathbf r}$ varying along the detector
surface \cite{Glauber}. These atoms have two internal discrete levels
$|g\rangle$ and $|e\rangle$, which are resonant with the wavelength
$\lambda_1=2\pi/k_1$. The level $|e\rangle$ is monitored for population
at time intervals $\delta t$ which we will take to be sufficiently small
so that the atomic dynamics can be neglected during that time. The level
$|e\rangle$ has a width $\gamma$: for sufficiently large values of
$\gamma$ our model corresponds to a broadband detector, whereas for
small values it corresponds to a narrowband detector. The results will
be independent of the specific value of $\gamma$. We will concentrate on
a given atom $C$ of the detector coupled to the quantized
electromagnetic field, which in turn is coupled to atoms $A$ and $B$. We
will calculate the state in which those atoms are left when the atom $C$
is found in the state $|e\rangle$, and we will add incoherently the
contributions corresponding to different detection times and different
positions ${\mathbf r}$. In such a way we will be finally able to derive
an expression for the density operator of atoms $A$ and $B$ conditioned
to the observation of a click of the detector.

Using standard methods of Quantum Optics, one can trace out the
electromagnetic field and obtain a master equation for the atoms $A$,
$B$ and $C$:
\be
\label{master}
\frac{d}{dt}\rho = \left[ {\cal L}^C + \sum_{\alpha=A,B} \left({\cal L}^\alpha + 
{\cal S}^{\alpha,C} + {\cal J}^{\alpha,C} \right) \right] \rho,
\ee
where ${\cal L}^\alpha$ denotes the Liouvillian superoperator describing
the evolution of atom $\alpha$ alone, and 
\bma
\bea
{\cal S}^{\alpha,C}\rho &=& - i \frac{\tilde\gamma}{2} G\left(|
{\mathbf r}^\alpha-{\mathbf r}|\right)
\left( \sigma_{eg}^C \otimes \sigma_{12}^\alpha + 
\sigma_{ge}^C \otimes \sigma_{21}^\alpha \right) \rho + H. c., \\
{\cal J}^{\alpha,C} \rho &=& \tilde\gamma \int\frac{d\Omega}{4\pi} 
e^{-i{\mathbf k}(\Omega)\cdot{\mathbf r}^\alpha} \sigma_{12}^\alpha
\rho \sigma_{eg}^C e^{i{\mathbf k}(\Omega)\cdot{\mathbf r}} + H. c. 
\,,
\eea
\ema
with $\sigma^\alpha_{ij}=|i\rangle_\alpha\langle j|$ (superscripts
indicate the atom, whereas subscripts indicate the states). Here and
in the following we will use the symbol $\otimes$ (tensor product)
whenever we feel that it clarifies the corresponding expression. The vectors
${\mathbf r}^A$ and ${\mathbf r}^B$ are the position operators of the
atoms $A$ and $B$, while the vector ${\mathbf r}$ is treated as a
c--number. The presence of the factor $G({\mathbf
r})=-\exp(ik_{1}|{\mathbf r}|)/(k_{1}|{\mathbf r}|)$ is due to the
dipole--dipole interaction (real part) and reabsorption (imaginary part)
between atoms $A,B$ and $C$, $\tilde\gamma$ giving the typical strength
of this interaction. These two terms give rise to the excitation of atom
$C$ via a photon absorption from atoms $A$ and/or $B$, which leads to a
detection event. We have assumed $k(d^2+D^2)^{1/2} \gg 1$, so that only
the far--field part contributes to the dipole--dipole interaction.

The Liouvillian action on atom $C$ (detector) is given by
\be
{\cal L}^C\rho = -\frac{\gamma}{2} (\sigma_{ee}^C\rho + \rho \sigma_{ee}^C) +
\gamma \sigma_{ge}^C \rho \sigma_{eg}^C.
\ee
In the absence of laser excitation, we have ($\alpha=A,B$)
\bea
{\cal L}^\alpha &=& \frac{1}{i\hbar} \left[H_{\rm tp}^\alpha,\rho\right] 
  - \frac{\Gamma}{2} \left(\sigma_{22}^\alpha \rho + \rho \sigma_{22}^\alpha \right) \\
&& + \Gamma_0 \int \frac{d\Omega}{4\pi} 
  N_0(\Omega) e^{-i{\mathbf k}_0(\Omega)\cdot{\mathbf r}^\alpha} \sigma_{02}^\alpha 
  \rho \sigma_{20}^\alpha e^{i{\mathbf k}_0(\Omega)\cdot{\mathbf r}^\alpha} \nonumber\\
&& + \Gamma_1 \int \frac{d\Omega}{4\pi} 
  N_1(\Omega) e^{-i{\mathbf k}_1(\Omega)\cdot{\mathbf r}^\alpha} \sigma_{12}^\alpha 
\rho \sigma_{21}^\alpha e^{i{\mathbf k}_1(\Omega)\cdot{\mathbf r}^\alpha}. \nonumber
\eea
Here, $H_{\rm tp}$ is the Hamiltonian describing the motion of an atom
in an isotropic harmonic potential of frequency $\nu$, and $N_0$ and $N_1$
describe the dipole emission pattern corresponding to transitions 
$|2\rangle\rightarrow|0\rangle$ and $|2\rangle\rightarrow|1\rangle$,
respectively. 

The master equation (\ref{master}) can be solved formally as 
\bea
\label{eq:rho}
\rho(t) &=& e^{({\cal L}^A + {\cal L}^B + {\cal L}^C)(t-t_0)}\rho(t_0) \\
&& + \int_{t_0}^t d\tau e^{({\cal L}^A + {\cal L}^B + {\cal L}^C)(t-\tau)}
\left[ {\cal S}^{A,C} + {\cal J}^{A,C} + {\cal S}^{B,C} + {\cal J}^{B,C}
\right] \rho(\tau).\nonumber
\eea
This integral equation can be iterated to obtain a formal expansion in
terms of ${\cal S}$ and ${\cal J}$. Since each of these terms scales as
$1/k(d^2+D^2)^{1/2}\ll 1$, we can stop at the first non--vanishing order
of the equation. The even terms of the expansion correspond to physical
processes in which excitations (photons) are interchanged between atoms
$A$ and $C$ (or $B$ and $C$). We have not included in Eq.\ (\ref{master}) 
the (dipole--dipole) interactions between atoms $A$ and $B$ which would
give rise to processes describing photon exchange, because they
correspond to a very small correction of the order of $1/kd\ll 1$ to the
final result. Note that we should only consider the case in which atom
$C$ is detected in $|e\rangle$, which can only occur if a photon coming
from $A$ or $B$ is absorbed; that is, the first non--vanishing process
in our expansion will correspond to the emission of a photon from atom
$A$ or $B$ subsequently absorbed by atom $C$. This will give a
contribution of the order $1/k^2(d^2+D^2)$. Processes in which more than
one photon are interchanged between atoms $A$ (or $B$) and $C$, or in
which (apart from the photon absorbed by $C$) other photons are
interchanged between atoms $A$ and $B$ would give higher order
contributions, at least of the order of $1/k^4(d^2+D^2)^{2}$ or
$1/k^4(d^2+D^2)d^2$, respectively.

\subsection{Initial state of atoms $A$ and $B$: Laser interaction}

So far, we have ignored the initial state of atoms $A$ and $B$. Let us
assume that they are driven by a very short laser pulse of duration
$t_{\rm las}\ll \Gamma^{-1},\nu^{-1}$. The state of atom $\alpha$ after
the interaction is
\be
\label{eq:rho(0)}
\tilde \rho_\alpha(0) = e^{-i h_{\rm las}^\alpha} \rho^\alpha(0) e^{i h_{\rm las}^\alpha},
\ee
where $\rho^\alpha(0)=\sigma_{00}^\alpha \otimes \rho_{\rm tp}^\alpha(0)$, with
\be
\label{mot}
\rho_{\rm tp}^\alpha(0) \propto \exp(-H_{\rm tp}^\alpha/k_BT)
\ee
being the initial motional state corresponding to a thermal distribution at 
temperature $T$ in the trapping potential, and $e^{-i h_{\rm las}^\alpha}$
acts in the subspace span$\{|0\rangle_\alpha,|2\rangle_\alpha\}$ as
\be
e^{-i h_{\rm las}^\alpha} = \cos(\theta_{\rm las})- i \sin(\theta_{\rm las})
\left[ \sigma_{20}^\alpha e^{i{\mathbf k}^\alpha \cdot{\mathbf r}^\alpha} + H. c.
\right].
\ee
Here, $\theta_{\rm las}$ is the rotation angle due to the laser interaction and
${\mathbf k}^\alpha$ the laser wavevector acting on atom $\alpha$. 

According to these equations, the effect of the laser on each of the
atoms is twofold: on one hand, it excites a superposition of the
internal states $|0\rangle$ and $|2\rangle$; on the other hand, it gives
a kick to the atom. The coefficient of the superposition $\theta_{\rm
las}$ can be easily varied by changing the laser intensity/duration.

\subsection{Detection}

We will use the following model for the detection \cite{Cohen}. The
initial state of the atom detector is $|g\rangle$. The evolution time is
divided in time steps $t_1,t_2, \ldots,t_n,\ldots$ of duration $\delta
t\ll \Gamma^{-1},\nu^{-1}$. After each time interval $\delta t$, the
internal state of atom $C$ is measured and the state of the whole system
is projected onto $|g\rangle$ or $|e\rangle$ depending on the outcome.
Let us consider the case in which the detection at time
$t_1,t_2,\ldots,t_n$ has yielded the outcome $|g\rangle$, and the
detection at time $t_{n+1}$ has yielded $|e\rangle$. To lowest order in
our expansion, the unnormalized state of atoms $A$ and $B$ at time
$t\rightarrow\infty$ once we have made the corresponding projections
will be
\be
\rho_n = K \lim_{t\to\infty} e^{({\cal L}^A + {\cal L}^B )(t-t_n)}R(t_n),
\ee
where $K$ is a constant that only depends on $\gamma,\tilde\gamma$ and $\delta t$, and
\bea
\label{eq:R}
R(t) &=& G \left({\mathbf r}^A-{\mathbf r}\right) \sigma^A_{12} \rho^A(t) \sigma^A_{21} 
  G\left({\mathbf r}^A-{\mathbf r}\right)^\dagger  \otimes \rho^B(t)\\
&&+ G\left({\mathbf r}^A-{\mathbf r}\right) \sigma_{12}^A \rho^A(t)\otimes
  \rho^B(t) \sigma^B_{21} G\left({\mathbf r}^A-{\mathbf r}\right)^\dagger \nonumber\\
&&  + \hbox{ same with} \; A \leftrightarrow B,\nonumber
\eea
with $\rho^\alpha(t)= e^{{\cal L}^\alpha t} \rho^\alpha(0)$.
This expression along with other intermediate results are calculated in the Appendix.

Since we do not know a priori at which time the detection will take place, we
have to perform the sum over all the operators $\rho(t_n)$. This sum can be transformed
into an integral given the fact that $\delta t$ is smaller than any dynamical
parameter corresponding to the evolution of atoms $A$ and $B$. Moreover, we also
have to integrate to all positions ${\mathbf r}$ corresponding to the detector; that is,
to all positions of atom $C$. By doing so, we are adding {\em incoherently}
all the contributions coming from detections at different points of the detector.
Finally, we have to trace over the motional states of atoms $A$ and $B$. The result,
properly normalized, will give the averaged density operator provided the detector
has performed a click (i.e., detected one photon).

\section{Results}

\subsection{Density operator and fidelity}

As it is shown in the Appendix, the reduced density operator describing
the internal state of atoms $A$ and $B$ in the case of detection can be
written as the sum of two contributions
\be
\label{rhoAB}
\rho^{AB} = \frac{R_1 + R_2}{{\rm tr}(R_1 + R_2)},
\ee
where 
\bma
\bea
R_1 &=& \cos^2(\theta_{\rm las}) \sin^2(\theta_{\rm las}) \nonumber \\
& & \left[ M^{A,A} |1,0\rangle\langle 1,0| + M^{B,B} |0,1\rangle\langle 0,1| + 
  M^{A,B} |1,0\rangle\langle 0,1| + M^{B,A} |0,1\rangle\langle 1,0|\right]\\
R_2 &=& \sin^4(\theta_{\rm las}) \nonumber \\
& & \left[ \frac{\Gamma_0}{\Gamma} M^{A,A} |1,0\rangle\langle 1,0| 
  + \frac{\Gamma_0}{\Gamma} M^{B,B} |0,1\rangle\langle 0,1| + 
  \frac{\Gamma_1}{\Gamma} (M^{A,A}+M^{B,B}) |1,1\rangle\langle 1,1|\right].
\eea
\ema
Here, we have defined
\be
\label{MMM}
M^{\alpha,\beta} = \int_S d{\mathbf r} \int_0^\infty dt\Gamma e^{-\Gamma t}
  {\rm tr}_{\rm tp} \left \{ G\left({\mathbf r}^\alpha(t) -{\mathbf r}\right)
  e^{i{\mathbf k}^\alpha\cdot {\mathbf r}^\alpha(0)}
  \rho_{\rm tp}^A(0)\rho_{\rm tp}^B(0) e^{-i{\mathbf k}^\beta\cdot 
  {\mathbf r}^\beta(0)} 
  G\left({\mathbf r}^\beta(t) -{\mathbf r}\right)^\dagger \right \}\,,
\ee
where the first integral is extended to the detector surface, the trace
is taken over the motional states of both atoms, and $\rho_{\rm
tp}^{A,B}(0)$ denote the initial motional states (\ref{mot}). The
time--dependent operators ${\mathbf r}^\alpha(t)=\exp(iH^\alpha_{\rm tp}t)
{\mathbf r}^\alpha \exp(-iH_{\rm tp}^\alpha t)$ are defined in the
interaction picture with respect to the harmonic potential. 

The interpretation of Eq.\ (\ref{rhoAB}) is very simple. 
The term $R_1$ comes from processes in which only one atom is excited by
the laser pulses and the subsequent photon emission is captured at the
detector. This can be easily understood if one writes such a term as 
\be
\label{R1}
R_1 = \int_S d{\mathbf r} \int_0^{\infty} dt\Gamma e^{-\Gamma t} {\rm tr}_{\rm tp} 
\left \{|\psi(t)\rangle\langle\psi(t)| \right \}\,,
\ee
with 
\be
|\psi(t)\rangle_{A,B} = G\left({\mathbf r}^A(t) -{\mathbf r}\right) 
  e^{i{\mathbf k}^A\cdot {\mathbf r}^A(0)}|1,0\rangle_{A,B}
  + G\left({\mathbf r}_B(t) -{\mathbf r}\right) 
  e^{i{\mathbf k}^B\cdot {\mathbf r}^B(0)}|0,1\rangle_{A,B}.
\ee
The state $|\psi(t)\rangle$ is the superposition of two states. The
first one comes from the process in which at time zero the laser excites
atom $A$, including the corresponding recoil; then, at time $t$ the atom
emits a photon which is detected by the atomic detector at position
${\mathbf r}$. The factor $G\left({\mathbf r}^A(t) -{\mathbf r} \right)$ 
includes the phase acquired during the propagation from the position of atom 
$A$ to the detector as well as the attenuation of the probability of reaching
the detector which is inversely proportional to the distance traveled
(a solid angle factor). The second term has the same contribution but
for the process in which atom $B$ is excited. Since we do not take into
account the exact time at which the photon is detected, we have to
multiply $|\psi(t)\rangle\langle\psi(t)|$ by the probability density
that the photon is emitted at time $t$, proportional to
$e^{-\Gamma t}$, and integrate over time. On the other hand, since we do
not know the point at the detector where the photon arrives, we have 
also to integrate the resulting expression over the detector surface, 
resulting in Eq.\ (\ref{R1}). Notice that retardation effects are not 
included in our formulation. They can be simply incorporated to this formula by
 changing $t\rightarrow t-|{\mathbf r}^{A,B}(t) -{\mathbf r}|/c$. Since here
${\mathbf r}^{A,B}$ and ${\mathbf r}$ vary over very small distances (size of the
atomic wavepackets and detector size, respectively), the result will not
be affected by retardation effects. On the other hand, expanding the
term $R_2$ in a similar way as Eq.\ (\ref{R1}) one can readily see that it
comes from the process in which both atoms are excited by laser pulses,
one photon emission is detected and the other not. The terms
proportional to $\Gamma_0$ correspond to the case in which the
undetected photon is emitted in the transition $|2\rangle\to|0\rangle$,
whereas the ones with $\Gamma_1$ correspond to the
$|2\rangle\to|1\rangle$ transition. 

With these expressions, we can easily calculate the fidelity (\ref{fidel}) as
\be
F= \frac{1}{2} \cos^2(\theta_{\rm las}) \left[ 1 + 
   \frac{M^{A,B}e^{i\phi}+M^{B,A}e^{-i\phi}}{M^{A,A}+M^{B,B}}\right]
  + \frac{\Gamma_0}{2\Gamma} \sin^2(\theta_{\rm las})\,,
\ee
where $\phi$ is the phase introduced in Eq.~(\ref{Bell}).
Given the fact that the size of the atom wavepackets is much smaller
than $D$, we can further simplify these expressions. First, we write
${\mathbf r}^\alpha={\mathbf r}^\alpha_0 + {\mathbf s}^\alpha$ with 
$|{\mathbf r}_\alpha^0-{\mathbf r}|\gg {\overline s}^\alpha$, 
the typical value taken by the operator ${\mathbf s}^\alpha$ (of the 
order of the size of the atomic wavepacket). Then, we expand
\be
\label{GGG}
G\left({\mathbf r}^{\alpha(t)} -{\mathbf r}\right) e^{i{\mathbf k}^\alpha\cdot 
{\mathbf r}^\alpha(0)} \simeq
  - \frac{e^{i({\mathbf k}^\alpha\cdot {\mathbf r}^\alpha_0 + k_1 |
  {\mathbf r}^\alpha_0-{\mathbf r}|)}}
  {k_1|{\mathbf r}^\alpha_0-{\mathbf r}|}
  e^{-i {\mathbf k}_1^\alpha {\mathbf s}^\alpha(t)} e^{i {\mathbf k}^\alpha 
  {\mathbf s}^\alpha(0)}\,, 
\ee
where ${\mathbf k}_1^\alpha$ is a vector of modulus $k_1=2\pi/\lambda_1$ and
direction given by ${\mathbf r}-{\mathbf r}^\alpha_0$. The integrals extended to
the detector in Eq.\ (\ref{MMM}) can then be performed using standard methods
of classical optics (substituting ${\mathbf r}$ by ${\mathbf r}_0$ in the
denominator of Eq.\ (\ref{GGG}), and expanding ${\mathbf r}$ around 
${\mathbf r}_0$ in the exponential for $M^{A,B}$ and $M^{B,A}$). Taking for simplicity
$|{\mathbf r}^A_0-{\mathbf r}_0|= |{\mathbf r}^B_0-{\mathbf r}_0|=(d^2+D^2)^{1/2}$ 
we find $M^{A,A}=M^{B,B}= L_xL_y/(d^2+D^2)$ and 
\be
M^{A,B} =(M^{B,A})^\ast= M^{A,A} e^{i({\mathbf k}^A\cdot {\mathbf r}^A_0 - 
{\mathbf k}^B\cdot {\mathbf r}^B_0)}
F_{\rm geo} F_{\rm dyn} \,,
\ee
where
\bma
\bea
F_{\rm geo} &=& \frac{1}{L_xL_y} \int_{-L_x/2}^{L_x/2} dx 
  \int_{-L_y/2}^{L_y/2} dy e^{-i k_1  x d/(d^2+D^2)^{1/2}} \\
F_{\rm dyn} &=& \Gamma \int_0^\infty dt e^{-\Gamma t} 
  {\rm tr}_{\rm tp}\left[ e^{-i{\mathbf k}_1^A \cdot {\mathbf s}^A(t)} 
   e^{-i{\mathbf k}^A \cdot {\mathbf s}^A(0)} \rho_{\rm tp}^A(0) \right]
  {\rm tr}_{\rm tp}\left[ \rho_{\rm tp}^B(0) e^{i{\mathbf k}^B \cdot {\mathbf s}^B(0)} 
  e^{-i{\mathbf k}_1^B \cdot {\mathbf s}^B(t)}   \right].
\eea
\ema
Evidently, $F_{\rm geo}$ coincides with Eq.\ (\ref{Fgeo}). On the other
hand, denoting by $\chi$ the angle between ${\mathbf k}_1^A$ and
${\mathbf k}^A$, which for simplicity we take to be equal to the angle
between ${\mathbf k}_1^B$ and ${\mathbf k}^B$, we obtain Eq.\
(\ref{Fdyn}). By further choosing $\phi=- ({\mathbf k}_A\cdot {\mathbf
r}_A^0 - {\mathbf k}_B\cdot {\mathbf r}_B^0)$ we obtain 
\be
F= \frac{1}{2} \cos^2(\theta_{\rm las}) \left[1+F_{\rm geo} F_{\rm dyn} \right]
+ \frac{\Gamma_0}{2\Gamma} \sin^2(\theta_{\rm las}).
\ee
Taking the worst case $\Gamma_0=0$, we finally arrive at Eq.\  (\ref{Fid}).

\subsection{Detection probability}

In order to derive an expression for the detection probability we just
have to combine geometrical considerations with the detection efficiency $\eta_D$
and the excitation probability. The probability of detection of a 
emitted photon is given by 
\be
P_0 = \eta_D \frac{D}{(d^2+D^2)^{1/2}} \frac{L_xL_y}{4\pi (d^2 + D^2)}
\,,
\ee
being $\eta_{D}$ the quantum efficiency of the photon detector. 
The first quotient in the expression is the cosine of the angle between the
vector connecting the atoms and the center of the detector with a vector 
perpendicular to its surface. The second one is the solid angle extended
by the detector from the atoms position. The probability that one and only
one atom is excited and the corresponding emitted photon detected is $P_0
2\sin^2(\theta_{\rm las})\cos^2(\theta_{\rm las})$. The probability that
both atoms are excited and one of the emitted photons is detected is 
$P_0 2\sin^4(\theta_{\rm las})$ (we neglect the process in which both photons
go to the detector). Thus, the desired probability is 
\be
P_{\rm det}= \sin^2(\theta_{\rm las}) \eta_D \frac{DL_xL_y}{2\pi(d^2+D^2)^{3/2}}. 
\ee
The maximum probability occurs for $D=d/\sqrt{2}$.

\section{Discussion}

As shown in the previous sections, using our proposal, one can create
states close to the maximally entangled state (\ref{Bell}). 
A typical test to determine whether one has succeeded or not, such as 
searching for violations of the CHSH inequalities \cite{1}, would require the
repetition of the experiment several times, and different measurements
on the internal atomic states. A positive result would occur if $F\agt
0.79$, something imposing restrictive conditions on the 
parameters of the experimental setup. 

To create an entangled state of high fidelity
the following conditions are required (cf. Eq.\ (\ref{Fid}) and
Eq.\ (\ref{Factors})): first, $\epsilon_1\equiv\sin^2(\theta_{\rm las})\ll
1$; second, $\epsilon_2\equiv dL_x/[2\lambda_1(d^2+D^2)^{1/2}] \ll 1$;
third, either $\epsilon_3\equiv 2\eta^2\coth(\hbar\nu/2k_BT) (\nu/\Gamma)^2 \ll
1$ (weak confinement) or $\epsilon_3\equiv 2\eta^2\coth(\hbar\nu/2k_BT) \ll
1$ (strong confinement). The first two conditions immediately imply a detection
probability $P_{\rm det} \ll 1$. In terms of these parameters we have
\bma
\bea
F &\simeq& 1- \frac{1}{2} \left[\epsilon_1+\frac{\epsilon_2^{2}}{6}+\epsilon_3\right],\\
P_{\rm det} &=& \frac{4}{\pi}\eta_D \epsilon_1 \epsilon_2^{3} 
\frac{L_{y}}{L_{x}}\frac{D}{L_{x}}
\left ( \frac{\lambda_{1}}{d} \right )^{3}.
\label{Pdet}
\eea
\ema
Choosing a favorable case such as $\epsilon_1=0.1$, $\epsilon_2=0.5$,
$\epsilon_3=0.1$, still gives rise to a fidelity $F > 0.8$ (i.e.,
Bell inequalities are still violated). Let us analyze for this case how 
``distant'' the atoms can be for sensible values of the parameters.
Rewriting $\epsilon_{2}$ as
\be
\epsilon_{2} = \frac{1}{2}\left [\frac{\lambda_{1}}{d} 
\sqrt{\left (\frac{d}{L_{x}} \right )^{2}+
\left ( \frac{D}{L_{x}} \right )^{2}} \right ]^{-1} \,,
\ee

\noindent
a value 0.5 impose $D/L_{x} \simeq 50$, assuming it is not possible 
$L_{x}\ll d \simeq D$ that would minimize $\epsilon_{2}$ while 
maximizing $P_{{\rm det}}$. Substituting in Eq.\ (\ref{Pdet}) 
\be
P_{{\rm det}} = 0.8 \eta_{D}\frac{L_{y}}{L_{x}} 
\left (\frac{\lambda_{1}}{d} \right )^{3} \,.
\ee
Considering an experiment is performed every $10^{-4}$ seconds (as it is
typically the case with trapped ions) a $P_{{\rm det}}=10^{-4}$ would
correspond to a detection per second. Then, with $L_{y}/L_{x}=30$ and a
$50\%$ efficiency, a separation of 100 wavelengths is possible. Notice
that the observation times cannot be increased arbitrarily for the
deleterious effect caused by dark counts occurring at the detector
increases consequently. 

Still, we need to asses the feasibility of $\epsilon_{3}=0.1$ or 
equivalently of $F_{{\rm dyn}}=0.9$. In doing so, we will define a new 
parameter, i.e.,
\be
\label{eq:etaI}
\eta_{I} \equiv k_{1} \sqrt{\frac{\hbar}{2  m \Gamma}} \,,
\ee

\noindent
so that $\eta^2 = \eta_{I}^2 (\Gamma/\nu)$. The new parameter (a 
redefinition of the Lamb-Dicke parameter with $\Gamma$ replacing 
$\nu$) allow us to study the behavior of $F_{{\rm dyn}}$ with 
respect to $\nu/\Gamma$. Once an atom and transition are chosen,
$\eta_{I}$ is fixed. Then, different values of $\nu/\Gamma$ 
corresponds to different designs of the trap for the chosen atom and 
transition. In the weak confinement limit, for fixed $\eta_{I}$, 
$\epsilon_{3} \sim \nu/\Gamma$ (just substitute 
Eq.\ (\ref{eq:etaI}) in Eq.\ (\ref{eq:Fapro})), whereas in the strong 
limit $\epsilon_{3} \sim \Gamma/\nu$. In both extremes, then, 
$F_{{dyn}}$ approaches one. However, for the former case 
Eq.\ (\ref{eq:Fapro}) is not valid for arbitrarily low values of 
$\nu/\Gamma$ unless $\cos(\chi) = 1$ strictly. Any finite value of 
$\chi$ implies $F_{{dyn}}=0$ at $\nu/\Gamma = 0$. Actually $\chi$ must 
be finite in order to avoid the laser light to impinge the detector, 
and therefore the best we can expect is a local maximum for $F_{{\rm 
dyn}}$ close to one. On the other hand, the strong confinement limit 
can be illusory for dipole transitions (needed to detection of the 
spontaneously emitted photon in a reasonable time). We are bounded, 
then, to treat $F_{{\rm dyn}}$ exactly. In Figures \ref{fg:Fdyn} the 
behavior of the dynamical factor with respect to $\nu/\Gamma$ is 
displayed for two values of $\eta_{I}$, namely, 0.05 (Figure {\bf a}) 
and 0.3 (Figure {\bf b}). The value $\eta_{I}=0.05$ corresponds 
approximately to the case of the NIST experiment \cite{5,6}.
The $\chi$ angle has been set to $8^{o}$, far larger than the 
minimum needed to avoid the laser light to impinge on the detector 
($0.8^{o}$ for $D/L_{x} \simeq 50$). In both figures the optimum case 
of sideband cooling reaching $T=0$ is compared with standard laser cooling 
at the Doppler limit and half the way to it. The maximum of 
$\nu/\Gamma$ is set to one, corresponding to the trap frequency equaling a 
dipole transition decay rate. The value $\eta_{I}=0.3$ 
represents in such a case a limit for Doppler cooling reaching 
$\epsilon_{3}=0.1$. From the curves shown, Doppler cooling is far 
enough for guaranteeing $F_{{\rm dyn}}$ with sensible values of 
$\nu/\Gamma$.  

The main problem which makes the detection probability small and
prevents the creation of a macroscopic distant entangled state is the 
geometrical factor. The  factor referring to the laser pulse area simply reduces 
by a factor of 10 the detection probability. 
In order to reduce the effects of the geometrical factor,
one can use lenses to collect photons emitted in different directions.
One could also couple the atoms to optical fibers, which would allow to
create entangled atoms over longer distances. In fact, one could 
embed the atoms in optical cavities, so that, with a high probability the
emitted photons, would go to the cavity mode, and then to a fiber 
coupled to it. The extent to which this can be performed in practice
depends on (near--) future developments in cavity--QED.

One can easily generalize the scheme proposed here to the case of more
atoms. For example, one can take $N$ atoms, excite all of them weakly using
a short laser pulse, and wait for a photodetection. In the ideal case,
a state
\be
|\psi\rangle = \frac{1}{\sqrt{N}} (e^{i\phi_1}|1,0,0,\ldots,0\rangle
  + e^{i\phi_2}|0,1,0,\ldots,0\rangle + \ldots +
  e^{i\phi_N}|0,0,0,\ldots,1\rangle)
\ee
would be created. By using more photodetectors and observing more detection
events one could create more general entangled states, although with
a decreasing probability of success.

\acknowledgments

C. Cabrillo acknowledges hospitality at University of Innsbruck. This work was
supported in part by the Acciones Integradas No.\ HU/997-0030 
(Spain-Austria), grants No.\ TIC95-0563-C05-03, 
No.\ PB96-00819, CICYT (Spain) and Comunidad de Madrid 06T/039/96 
(Spain), the FWF (Austrian Science Foundation) and by the TMR network 
ERB--FMRX--CT96--0087.

\appendix
\section*{Calculation of $\rho^{AB}$}

Let us denote by ${\cal S}(t)$ the free evolution operator, i.e.,
\be
{\cal S}(t) = e^{({\cal L}^A + {\cal L}^B + {\cal L}^C) t} \,.
\ee

\noindent
Then, iterating twice, Eq.\ (\ref{eq:rho}) results in
\bea
\label{eq:iterho}
\rho(t) & = & {\cal S}(t-t_{0})\rho(t_0) + 
\int_{t_0}^t d\tau \,{\cal S}(t-\tau)
\left[ ({\cal S}^{A,C} + {\cal J}^{A,C}){\cal S}(\tau)\rho(t_{0}) +
A \leftrightarrow B \right]  + \nonumber \\
&& \int_{t_0}^{t} d\tau \int_{t_0}^{\tau} d\tau^{\prime} {\cal S}(t-\tau) 
\left[({\cal S}^{A,C} + {\cal J}^{A,C}){\cal S}(t-\tau^{\prime}) 
\left\{({\cal S}^{A,C} + {\cal J}^{A,C}){\cal 
S}(\tau^{\prime})\rho(t_{0})+
A \leftrightarrow B \right\}\right] + \nonumber \\
&& A \leftrightarrow B + O(({\cal S}^{\alpha,C})^{3})\,.
\eea

\noindent
We are here interested only in its projection onto the detector atom 
excited state, i.e., in $\langle e|\rho(t) |e 
\rangle $. The free evolution of the detector atom is governed by
\bea
e^{{\cal L}^C t}\sigma_{gg}^{C} &=&\sigma_{gg}^{C} \,, \nonumber \\
e^{{\cal L}^C t}\sigma_{eg}^{C} &=&
e^{-t \gamma/2 } \sigma_{eg}^{C} \,, \nonumber \\
e^{{\cal L}^C t}\sigma_{ge}^{C} &=&
e^{-t \gamma/2 } \sigma_{ge}^{C} \,, \nonumber \\
e^{{\cal L}^C t}\sigma_{ee}^{C} &=& 
e^{- t\gamma} \sigma_{ee}^{C} \,, \nonumber
\eea
and it is simply enough to be operated out of $\langle e |\rho(t) |e 
\rangle$ given the initial state $\rho(t_{0}) = 
\tilde{\rho}^{A}(t_{0})\otimes\tilde{\rho}^{B}(t_{0})\otimes \sigma_{gg}^{C} $. Thus, 
\be
\langle e| [{\cal S}(t-t_{0})\rho(t_{0})] | e \rangle = 
\langle e|{\cal S}(t-\tau)
\left[ ({\cal S}^{\alpha,C} + {\cal J}^{\alpha,C}){\cal S}(\tau)\rho(t_{0})
\right ] |e \rangle = 0 \,,
\ee
so that
\bea
\label{eq:prorho}
\lefteqn{\langle e|\rho(t) |e \rangle  = \int_{t_0}^{t} d\tau 
\int_{t_0}^{\tau} d\tau^{\prime} \, e^{-\gamma (t-\tau)} 
e^{-\frac{\gamma}{2} (\tau-\tau^{\prime})} \times} \nonumber \\
& &\langle e |\, e^{({\cal L}^{A}+{\cal 
L}^{B})(t-\tau)} \left [ {\cal S}^{A,C} e^{({\cal L}^{A}+{\cal 
L}^{B})(\tau-\tau^{\prime})} \left \{{\cal S}^{A,C} e^{({\cal L}^{A}+{\cal 
L}^{B})\tau^{\prime}} \rho(t_{0}) + {\cal S}^{B,C} e^{({\cal L}^{A}+{\cal 
L}^{B})\tau^{\prime}} \rho(t_{0}) \right \} \right ] |e \rangle + 
\nonumber \\ 
& & A \leftrightarrow B \; = \nonumber \\
& &\int_{t_0}^{t} d\tau 
\int_{t_0}^{\tau} d\tau^{\prime} \, e^{-\gamma (t-\tau)} 
e^{-\frac{\gamma}{2} (\tau-\tau^{\prime})} \times \nonumber \\
&& \left \{ 
\langle e | \left [ e^{{\cal L}^{B} t} \tilde{\rho}^{B}(t_{0}) 
\right ]\otimes \left [  e^{{\cal L}^{A}(t-\tau)} 
{\cal S}^{A,C} e^{{\cal L}^{A}(\tau-\tau^{\prime})}{\cal S}^{A,C} 
e^{{\cal L}^{A}\tau^{\prime}} \tilde{\rho}^{A}(t_{0}) \sigma_{gg}^{C}
\right ] |e \rangle + \nonumber \right . \\
&& \left .
\langle e | \left [ e^{{\cal L}^{A}(t-\tau)}{\cal S}^{A,C} 
e^{{\cal L}^{A}\tau}\tilde{\rho}^{A}(t_{0}) \right ] \otimes
\left [ e^{{\cal L}^{B}(t-\tau^{\prime})}{\cal S}^{B,C} 
e^{{\cal L}^{B}\tau^{\prime}}\tilde{\rho}^{B}(t_{0}) \sigma_{gg}^{C} 
\right ] |e \rangle \right \} + \nonumber \\
& & A \leftrightarrow B \,. 
\eea

\noindent
As explained in the text during the measurement process, the detector 
atom is projected $n$ times onto the ground state before being 
projected onto the excited state at time $t_{n+1}$. Under the 
condition $\delta t \ll \Gamma^{-1}, \nu^{-1}$ the evolution 
Liouvillians inside Eq.\ (\ref{eq:prorho}) between $t_{n}$ and 
$t_{n+1}$ can be left constant so that
\be
\langle e|\rho | e \rangle  \propto  \int_{t_n}^{t_{n+1}} d\tau 
\int_{t_n}^{\tau} d\tau^{\prime} \, e^{-\gamma (t-\tau)} 
e^{-\frac{\gamma}{2} (\tau-\tau^{\prime})} \tilde{R}(t_{n}) \; = \;
\frac{2}{\gamma^{2}} \left [ 1 + e^{-\gamma \delta t} - 
2 e ^{-\gamma \delta t /2} \right ] \tilde{R}(t_{n}) \,, \nonumber
\ee
where
\bea
\tilde{R}(t_{n}) & = &   
\langle e | \left [ e^{{\cal L}^{B} t_{n}}\tilde{\rho}^{B}(t_{0}) 
\right ]\otimes \langle e | \left [
{\cal S}^{A,C}{\cal S}^{A,C} 
e^{{\cal L}^{A} t_{n}} \tilde{\rho}^{A}(t_{0}) \sigma_{gg}^{C} 
\right ] |e \rangle + \nonumber \\
& & \langle e | \left [ 
{\cal S}^{A,C} {\cal S}^{B,C} 
e^{({\cal L}^{A}+{\cal L}^{B})t_{n}}\rho(t_{0}) \right ] 
|e \rangle + \nonumber A \leftrightarrow B \,. 
\eea

\noindent
Substituting the definitions of ${\cal S}^{\alpha,C}$ in 
the previous equation, changing $t_{n}$ by $t$ and denoting $e^{{\cal 
L}^{\alpha} t} \tilde{\rho}^{\alpha}(0)$ by $\rho^{\alpha}(t)$ one arrives to 
an expression proportional to Eq.\ (\ref{eq:R}).

To proceed further we need to integrate the free evolution of the 
atoms given by
\be
\dot{\rho}^{\alpha} = -i [H_{tp}, \rho^{\alpha}] - 
\frac{\Gamma}{2}\sigma^{\alpha}_{22} \rho^{\alpha} -\frac{\Gamma}{2} 
\rho^{\alpha} \sigma^{\alpha}_{22} +
\Gamma_{0} \int d \Omega e^{-i{\mathbf k}_{0}(\Omega) 
{\mathbf r}^{\alpha}}
\sigma^{\alpha}_{02} \rho^{\alpha} \sigma^{\alpha}_{20}
e^{i{\mathbf k}_{0}(\Omega) {\mathbf  r}^{\alpha}} + 
0 \leftrightarrow 1 \,.
\ee
  
\noindent
In a frame rotating with the trap Liouvillian the solution results in
\bea 
\rho^{\alpha}(t) & = & e^{-\frac{\Gamma}{2}\sigma^{\alpha}_{22} t} 
\rho^{\alpha}(0) e^{-\frac{\Gamma}{2}\sigma^{\alpha}_{22} t}+
\nonumber \\
&& \Gamma_{0} \int_{0}^{t} d \tau e^{-\Gamma \tau}
\int d \Omega e^{-i{\mathbf k}_{0}(\Omega) {\mathbf r}^{\alpha}(\tau)}
\sigma^{\alpha}_{02} \rho^{\alpha}(0) \sigma^{\alpha}_{20}
e^{i{\mathbf k}_{0}(\Omega) {\mathbf  r}^{\alpha}(\tau)}+ 
0 \leftrightarrow 1 \,.
\eea

\noindent
Taking into the account the initial condition (\ref{eq:rho(0)}) we 
have
\bea
\label{eq:limit}
\lefteqn{\lim_{t\to\infty} e^{({\cal L}^A + {\cal L}^B )(t-\tau)}R(\tau)
= 
\left [ \cos^{2}(\theta_{las}) \sigma^{A}_{00}\otimes 
\rho^{A}_{tp}(0) +
 \right.}    \nonumber\\ 
&& \sin^{2}(\theta_{las})\sigma^{A}_{00} \Gamma_{0}\int_{0}^{\infty} d 
\tau^{\prime} e^{-\Gamma \tau^{\prime}} \int d \Omega 
e^{-i{\mathbf k}_{0}(\Omega) {\mathbf r}^{B}(\tau^{\prime})}
e^{i {\mathbf k}^{B}{\mathbf r}^{B}(0)} 
\rho^{B}_{{\rm tp}}(0)e^{-i {\mathbf k}^{B}{\mathbf r}^{B}(0)}
e^{i {\mathbf k}_{0}(\Omega){\mathbf r}^{B}(\tau^{\prime})}  +
\nonumber\\ 
&&\left . \sin^{2}(\theta_{las})\sigma^{A}_{11} \Gamma_{1}\int_{0}^{\infty} d 
\tau^{\prime} e^{-\Gamma \tau^{\prime}} \int d \Omega 
e^{-i{\mathbf k}_{1}(\Omega) {\mathbf r}^{B}(\tau^{\prime})}
e^{i {\mathbf k}^{B}{\mathbf r}^{B}(0)} \rho^{B}_{{\rm tp}}(0)
e^{-i {\mathbf k}^{B}{\mathbf r}^{B}(0)}
e^{i {\mathbf k}_{1}(\Omega) {\mathbf r}^{B}(\tau^{\prime})} \right ]
\otimes \nonumber \\
&& \sin^{2}(\theta_{las}) e^{-\Gamma \tau} \sigma^{A}_{11}
G({\mathbf r}^{A}(\tau)-{\mathbf r}) e^{i{\mathbf k}^{A} {\mathbf r}^{A}(0)}
\rho^{A}_{{\rm tp}}(0) e^{-i{\mathbf k}^{A}{\mathbf r}^{A}(0)}
G({\mathbf r}^{A}(\tau)-{\mathbf r})^{\dagger} + \nonumber \\
&& \sin^{2}(\theta_{las})\cos^{2}(\theta_{las})e^{-\Gamma \tau} 
\sigma_{10}^{A}\otimes\sigma_{01}^{B} G({\mathbf r}^{A}(\tau)-{\mathbf r}) 
\rho^{A}_{{\rm tp}}(0)\rho^{B}_{{\rm tp}}(0) 
G({\mathbf r}^{B}(\tau)-{\mathbf r})^{\dagger} + \nonumber \\
&& A \leftrightarrow B 
\eea

\noindent
Rearranging terms, Eq.\ (\ref{eq:limit}) can be decomposed as $ 
R_{1}(\tau)+R_{2}(\tau)$ with
\bea
R_{1}(\tau) & = & \sin^{2}(\theta_{las})\cos^{2}(\theta_{las}) 
e^{-\Gamma \tau} \times \nonumber \\
&&\left [
G({\mathbf r}^{A}(\tau)-{\mathbf r}) e^{i {\mathbf k}^{A} {\mathbf r}^{A}(0)}
|1,0 \rangle + G({\mathbf r}^{B}(\tau)-{\mathbf r}) e^{-i {\mathbf k}_{B} 
{\mathbf r}^{B}(0)} |0,1 \rangle 
\right ] 
\rho_{{\rm tp}}^{A}\rho_{{\rm tp}}^{B} \times \nonumber \\
& & \left [ 
G({\mathbf r}^{A}-{\mathbf r}) e^{i {\mathbf k}^{A} {\mathbf r}^{A}(0)}
|1,0 \rangle + G({\mathbf r}^{B}-{\mathbf r}) e^{-i {\mathbf k}^{B} 
{\mathbf r}^{B}(0)} |0,1 \rangle 
\right ]^{\dagger}
\\
R_{2}(\tau) & = & \sin^{4}(\theta_{las}) e^{-\Gamma \tau} \times 
\nonumber \\
&& \left \{
\sigma_{11}^{A}\otimes\sigma_{00}^{B} 
\Gamma_{0}\int_{0}^{\infty} d 
\tau^{\prime} e^{-\Gamma \tau^{\prime}} \int d \Omega 
e^{-i{\mathbf k}_{0}(\Omega) {\mathbf r}^{B}(\tau^{\prime})}
e^{i {\mathbf k}^{B}{\mathbf r}^{B}(0)} 
\rho^{B}_{{\rm tp}}(0)e^{-i {\mathbf k}^{B}
{\mathbf r}^{B}(0)}e^{i {\mathbf k}_{0}(\Omega){\mathbf r}^{B}(\tau^{\prime})} 
\otimes \right .
\nonumber \\
&&G({\mathbf r}^{A}(\tau)-{\mathbf r}) 
e^{i{\mathbf k}^{A} {\mathbf r}^{A}(0)} \rho^{A}_{{\rm tp}}(0) 
e^{-i{\mathbf k}^{A}{\mathbf r}^{A}(0)}G({\mathbf r}^{A}(\tau)-
{\mathbf r})^{\dagger}
\nonumber \\
&& \sigma_{11}^{A}\otimes\sigma_{11}^{B} 
\Gamma_{1}\int_{0}^{\infty} d 
\tau^{\prime} e^{-\Gamma \tau^{\prime}} \int d \Omega 
e^{-i{\mathbf k}_{1}(\Omega) {\mathbf r}^{B}(\tau^{\prime})}
e^{i {\mathbf k}^{B}{\mathbf r}^{B}(0)} 
\rho^{B}_{{\rm tp}}(0)e^{-i {\mathbf k}^{B}{\mathbf r}^{B}(0)}
e^{i {\mathbf k}_{1}(\Omega){\mathbf r}^{B}(\tau^{\prime})} \otimes
\nonumber \\
&&G({\mathbf r}^{A}(\tau)-{\mathbf r}) 
e^{i{\mathbf k}^{A} {\mathbf r}^{A}(0)} \rho^{A}_{{\rm tp}}(0) 
e^{-i{\mathbf k}^{A}{\mathbf r}^{A}(0)}
G({\mathbf r}^{A}(\tau)-{\mathbf r})^{\dagger} + 
\nonumber \\
&& \left. A \leftrightarrow B \right \} \,.
\eea

\noindent
Tracing over the motional states and using the cyclic property of the 
trace the exponential terms in $R_{2}(\tau)$ cancel out making 
the integral in $\tau^{\prime}$ trivial. Integrating ${\mathbf r}$ over 
the detector area and $\tau$ with a density 
$\Gamma e^{-\Gamma \tau}$, $R_{1}$ and $R_{2}$ are finally obtained.


\begin{figure}
\caption{Sketch of the experimental setup as well as of the internal 
level structure of the atoms corresponding to the proposed experiment.}
\label{fg:setup}
\end{figure}

\begin{figure}
\caption{The behavior of $F_{{\rm dyn}}$ as a function of 
$\nu/\Gamma$ for two different values of $\eta_{I}$ and three different 
temperatures. $T_{D}$ denotes the Doppler limit temperature, i.e., 
$T_{D} = \hbar \Gamma/2 k_{B}$.}
\label{fg:Fdyn}
\end{figure}

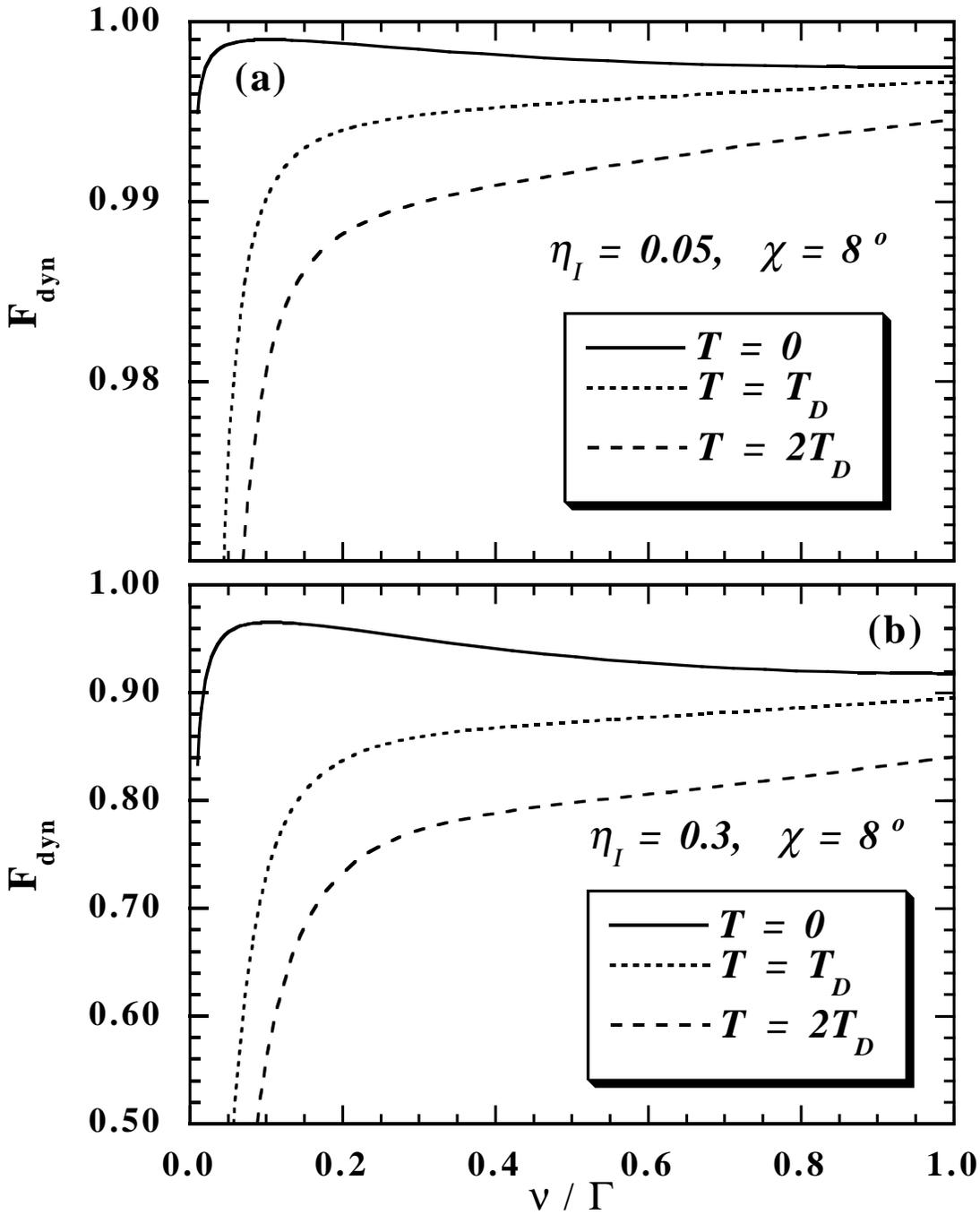
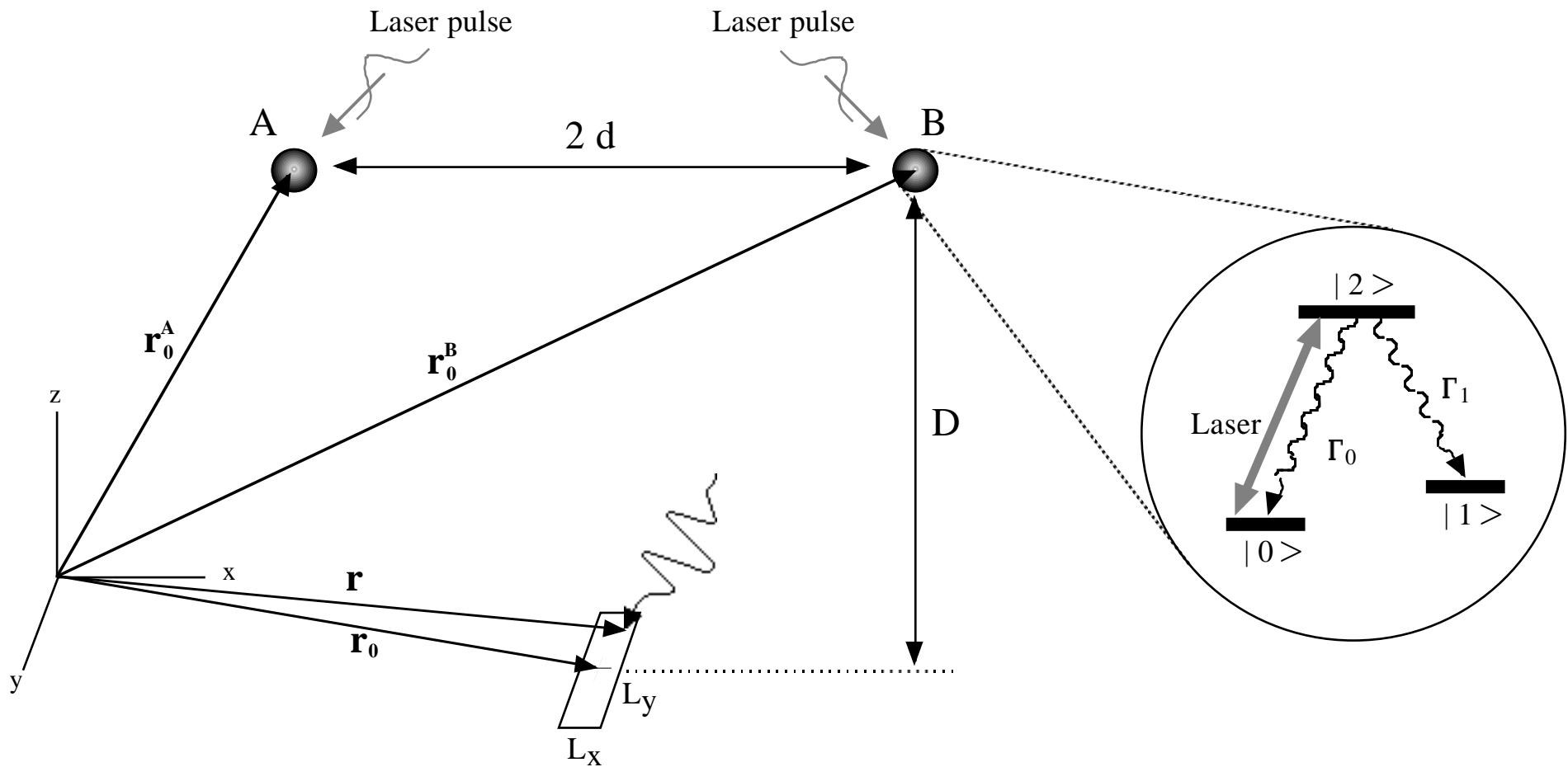
\end{document}